\documentclass[conference]{IEEEtran}
\IEEEoverridecommandlockouts
\usepackage{cite}
\usepackage{amsmath,amssymb,amsfonts}
\usepackage{algorithmic}
\usepackage{graphicx}
\usepackage{textcomp}
\usepackage{xcolor}

\usepackage{todonotes}
\usepackage{url}
\usepackage{siunitx}
\usepackage[left]{eurosym}

\def\BibTeX{{\rm B\kern-.05em{\sc i\kern-.025em b}\kern-.08em
    T\kern-.1667em\lower.7ex\hbox{E}\kern-.125emX}}

\makeatletter
\newcommand{\newlineauthors}{%
  \end{@IEEEauthorhalign}\hfill\mbox{}\par
  \mbox{}\hfill\begin{@IEEEauthorhalign}
}
\makeatother
    
\begin{document}

\title{Tool Wear Prediction in CNC Turning Operations using Ultrasonic Microphone Arrays and CNNs}

\author{
    \IEEEauthorblockN{
        Jan Steckel\IEEEauthorrefmark{1}\IEEEauthorrefmark{2}\IEEEauthorrefmark{5},
        Arne Aerts\IEEEauthorrefmark{4}, 
        Erik Verreycken\IEEEauthorrefmark{1}\IEEEauthorrefmark{2},
        Dennis Laurijssen\IEEEauthorrefmark{1}\IEEEauthorrefmark{2}, 
        Walter Daems\IEEEauthorrefmark{1}\IEEEauthorrefmark{2}
    }
    \IEEEauthorblockA{\IEEEauthorrefmark{1}Cosys-Lab, Faculty of Applied Engineering, University of Antwerp, Antwerp, Belgium}
    \IEEEauthorblockA{\IEEEauthorrefmark{2}Flanders Make Strategic Research Centre, Lommel, Belgium}
    \IEEEauthorblockA{\IEEEauthorrefmark{4}Service Group Materials, Department of Mechanical Engineering, University of Antwerp, Antwerp, Belgium}
    \IEEEauthorblockA{\IEEEauthorrefmark{5}jan.steckel@uantwerpen.be}
 }

\maketitle

\begin{abstract}
   This paper introduces a novel method for predicting tool wear in CNC turning operations, combining ultrasonic microphone arrays and convolutional neural networks (CNNs). High-frequency acoustic emissions between \SI{0}{\kilo\hertz} and \SI{60}{\kilo\hertz} are enhanced using beamforming techniques to improve the signal-to-noise ratio. The processed acoustic data is then analyzed by a CNN, which predicts the Remaining Useful Life (RUL) of cutting tools. Trained on data from 350 workpieces machined with a single carbide insert, the model can accurately predict the RUL of the carbide insert. Our results demonstrate the potential gained by integrating advanced ultrasonic sensors with deep learning for accurate predictive maintenance tasks in CNC machining.
\end{abstract}


\section{Introduction}
Predictive maintenance is the process of using sensor data to infer the state of a machine, and detect flaws or failures before they happen from that sensor data \cite{zontaPredictiveMaintenanceIndustry2020,zhangDatadrivenMethodsPredictive2019a}. Predictive maintenance is a rising trend in virtually all industry branches, ranging from bearing fault detection \cite{schwendemannSurveyMachinelearningTechniques2021}, gear wear detection \cite{raadnuiSpurGearWear2019} or piston wear in reciprocating machines \cite{shanbhagFailureMonitoringPredictive2021} amongst many other examples. A plethora of sensors is being applied for predictive maintenance, ranging from acoustics, vibration signals, thermal sensors, current sensors, and virtually any other sensing modality that can be imagined \cite{schwendemannSurveyMachinelearningTechniques2021, zontaPredictiveMaintenanceIndustry2020}. In previous work, we demonstrated bearing fault detection using ultrasound signals in the band of \SI{0}{\kilo\hertz} to \SI{60}{\kilo\hertz}, measured with a microphone array and subsequent beamforming techniques \cite{moto:c:irua:182053_vere_beam}. 

When machining complex parts with axis-symmetry, CNC lathes are the de-facto standard manufacturing technique. In a CNC lathe, the work piece is mounted to a spindle which is then being rotated, and cutting is performed by means of a stationary cutting tool. There is a wide variety of cutting tool geometry and materials, with carbides being the dominantly used material for fabricating these cutting tools. While the variety in geometry and material type is vast, one recurring and important feature of these tools is the sharpness. Indeed, the sharpness of the cutting tool often directly correlates to the finished quality of the workpiece \cite{gokkayaEffectsCuttingTool2007, sivaiahEffectSurfaceTexture2020}. A tool that is too dull will not cut adequately, leaving a poor surface finish, while a tool that is overly sharp is brittle and wears quickly. Manufacturers of CNC tools often specify a tool life parameter, expressed in hours, which indicates how long a tool can be utilized. However, these estimates are sometimes too conservative \cite{chanCNCCuttingToolsLife2022, gokkayaEffectsCuttingTool2007}, and therefore lead to premature replacement of still useful tools. Furthermore, some tools might fail prematurely, causing problems in automated manufacturing applications, which results in scrapped parts and lost revenue.

To combat the problem with tool wear inherent to CNC machining, various techniques for automated tool wear estimation have been devised \cite{chanCNCCuttingToolsLife2022, szecsiAutomaticCuttingtoolCondition1998}. Similar as to the general field of predictive maintenance, a wide range of sensing modalities is being used, such as consumed spindle power/current \cite{neefToolWearSurface2018}, vibration measurements \cite{arslanToolConditionMonitoring2016}, Doppler radar \cite{smithMethodDetectingTool2005} and acoustics \cite{liuToolWearMonitoring2019, salgadoApproachBasedCurrent2007, seemuangUsingSpindleNoise2016}. In many previous publications, combinations of these aforementioned sensor modalities are used to perform the tool wear or tool life prediction.

In this paper, we will focus on the use of ultrasonic acoustic signals in the range of \SI{0}{\kilo\hertz} to \SI{60}{\kilo\hertz} using an array sensor, similar to our previous work on bearing fault detection \cite{moto:c:irua:182053_vere_beam}. The sensor is based around our eRTIS ultrasonic sensor \cite{moto:c:irua:165188_kers_a} which allows the implementation of broadband spatial filters through beamforming. Using beamforming in high-noise scenarios such as CNC operations should improve the SNR of the signal, by filtering out unwanted noise sources. We will use our eRTIS sensor to measure the ultrasonic spectrogram acoustic signals generated by the cutting operation, and use beamforming to remove unwanted interference. Then, we will employ a convolutional neural network to perform Remaining Useful Life (RUL) prediction \cite{wangRemainingUsefulLife2020} of the cutting tool, using data obtained from 350 work pieces turned using a single carbide insert.

\begin{figure*}[t]
    \centering
    \includegraphics[width=\linewidth]{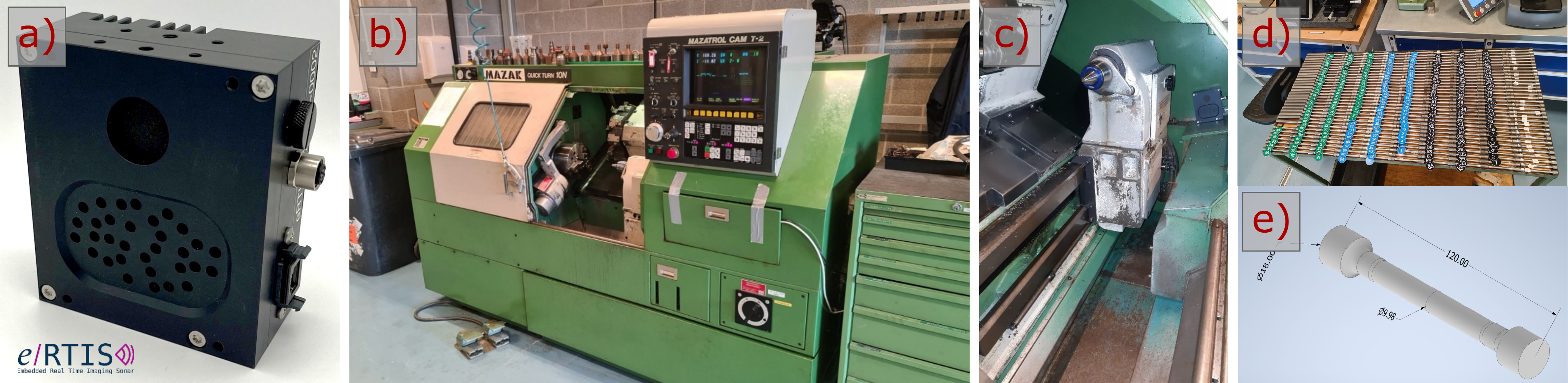}
    \caption{Overview of the hardware setup used in this paper. Panel a) shows the eRTIS sensor, an ultrasonic array sensor with 32 microphones arranged in a pseudorandom pattern. The sensor is waterproofed using a membrane material from the  Acoustic Protective Material line manufactured by W. L. Gore and Associates. Panel b) shows the CNC lathe being used, a Mazak Quickturn 10N, using Iscar GRIP 3015Y grade IC808 inserts. Panel c) shows the inside eRTIS sensor mounted in the lathe. Panel d) shows the turned materials, and e) shows the model of the piece that has been turned during this experiment.} 
    \label{fig:overview}
\end{figure*}

\section{Hardware Setup}
In this section we will provide some details on the experimental setup. We used an industrial CNC lathe (Mazak QT10N), which is shown in figure \ref{fig:overview} panel b). On this lathe, we turned a batch of tensile test specimen used in the mechanical engineering curriculum of our university until tooling insert failure (350 pieces). The model of this tensile test specimen can be seen in figure \ref{fig:overview}, panel e). We used two materials for making these test pieces, more specifically 1.1191 (C45) and 1.7225 (Chromoly), distributed as randomly as possible during the production run. We used a depth of cut of \SI{0.3}{\milli\meter}, a feed rate of \SI{0.5}{\milli\meter} per revolution and a surface speed of \SI{120}{\meter\per\minute}. As tooling insert we used an Iscar GRIP 3015Y in grade IC808.

As ultrasonic sensors we used two eRTIS sensors\cite{moto:c:irua:165188_kers_a}, which are ultrasonic microphone-array sensors with 32 microphones, sampled at \SI{450}{\kilo\hertz}. The individual sensors are synchronized using our pseudo-random 1-bit synchronization method described in  \cite{moto:c:irua:149170_laur_sync}. One eRTIS sensor was placed inside of the lathe, another outside right where the operator would stand. In addition, we recorded the spindle current using a Fluke i1000s current clamp, connected to a National Instruments NI-USB6363 DAQ, which sampled both the current signal as well as the synchronization signal at \SI{500}{\kilo\hertz}. The eRTIS sensors measured chunks of acoustic data of \SI{40}{\milli\second} at a rate of \SI{10}{\hertz} (i.e., the signal is sampled intermittently).

\begin{figure}[t]
    \centering
    \includegraphics[width=\linewidth]{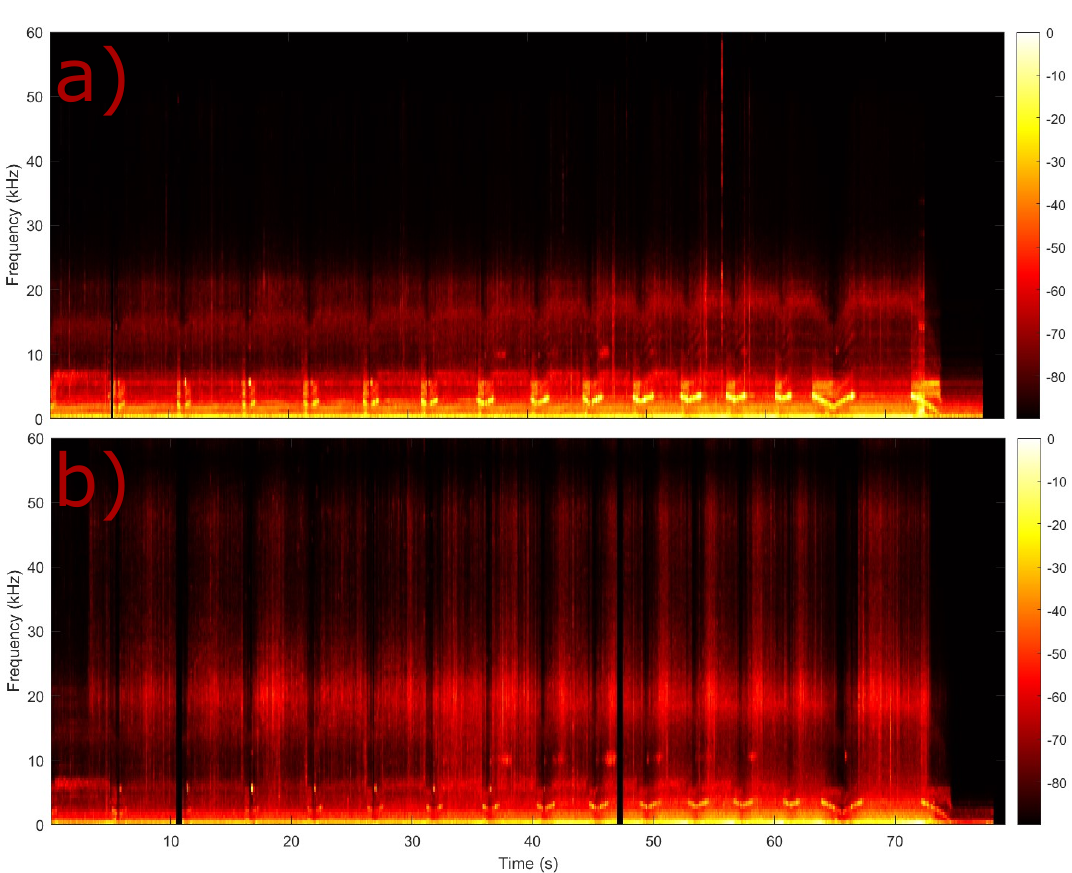}
    \caption{Spectrogram representation of the acoustic data recorded by the eRTIS sensors during a turning run. Panel a) shows the data of measured by the eRTIS sensor ouside of the machine, and panel b) shows the data recorded from inside the machine. In the low frequency parts, the signal is dominated by the motion sounds of the machine, and the high frequency parts of the spectrogram are generated by the cutting operations.} 
    \label{fig:spectros}
\end{figure}

\section{Signal Processing and Experimental Results}
In this paper, we treat each eRTIS sensor independently - i.e. there is no sensor fusion using multiple sensors. Each sensor measures 32 microphone signals $s_m(t)$. These microphone signals are passed through a beamformer $B_\psi$ pointing in direction $\psi$, which is the vector pointing from the sensor to the tool position. The beamformed signals are also filtered using a bandpass filter $h_{bp}(t)$, which is set to be a Butterworth filter of order 6 between \SI{0}{\kilo\hertz} and \SI{60}{\kilo\hertz}. The overall output of each datasample of \SI{40}{\milli\second} can then be calculated as:
\begin{equation}
    s_o(t) = h_{bp}(t) * \sum_{i=1}^{32} B_\psi \bigg ( s_m(t) \bigg )
\end{equation}
This output signal is then transformed to a magnitude spectrum $S_n(\omega)$ using a Welch Power Spectral Density estimator, using a window length of 1024 samples and Hamming weighting. In this spectrum $S_n(\omega)$,  the index $n$ represents the frame number of that data section (sampled at \SI{10}{\hertz}, and \SI{40}{\milli\second}). These spectra are all concatenated into a large matrix $S$, which then represents a spectrogram-like representation:
\begin{equation}
    S(\omega,n) = \bigg [ \begin{matrix} S_1(\omega) & S_2(\omega) & ... & S_N(\omega) \end{matrix} \bigg ]
\end{equation}
Examples of these spectrogram-like representations can be seen in figure \ref{fig:spectros}. Panel a) shows the spectrogram of a run recorded using an eRTIS outside of the lathe, and panel b) shows the spectrum recorded inside the machine. The main difference between the data recorded by these two sensors can be found in i) the low-frequency motion sounds, which are more clearly visible in the spectrum measured outside of the machine, and ii) the cutting noise, which is high in frequency and more prominent in the data in panel b). Both representations are shown on a logarithmic dB scale, normalized between 0dB and 90dB. One of these spectrograms is recorded for each of the 350 pieces that is being manufactured.

Finally, we implemented a convolutional neural network which solves the following regression task: estimating the work-piece number based on a acoustic spectrogram data (ie, a number between 1 and 350). The CNN is a standard feed-forward CNN with four convolutional layers, leaky ReLu  function, max pooling, layer normalization and dropout (10 percent). After the four convolutional layers the output gets flattened into a vector, and passed through two fully connected (FC) layers, where the first FC layer has an output dimension of 10 and a ReLu nonlinearity, and the last FC layer has a single output. We trained the neural network using the Adam optimizer, with a learning rate of 0.01 for a maximum of 100 epochs, and used the best validation loss checkpoint as the final network. We used data augmentation (shifting in the time dimension and additive noise), and split the augmented dataset randomly into a training set (75\%), validation (10\%) and test (15\%) sets. In the ideal case, we would perform this split before augmentation (or even use a complete second run), which would required a more extensive dataset, and that was not available at the time of writing. Training of the CNN on a single NVidia RTX4090 took around one hour, including data pre processing.

\begin{figure}[t]
    \centering
    \includegraphics[width=\linewidth]{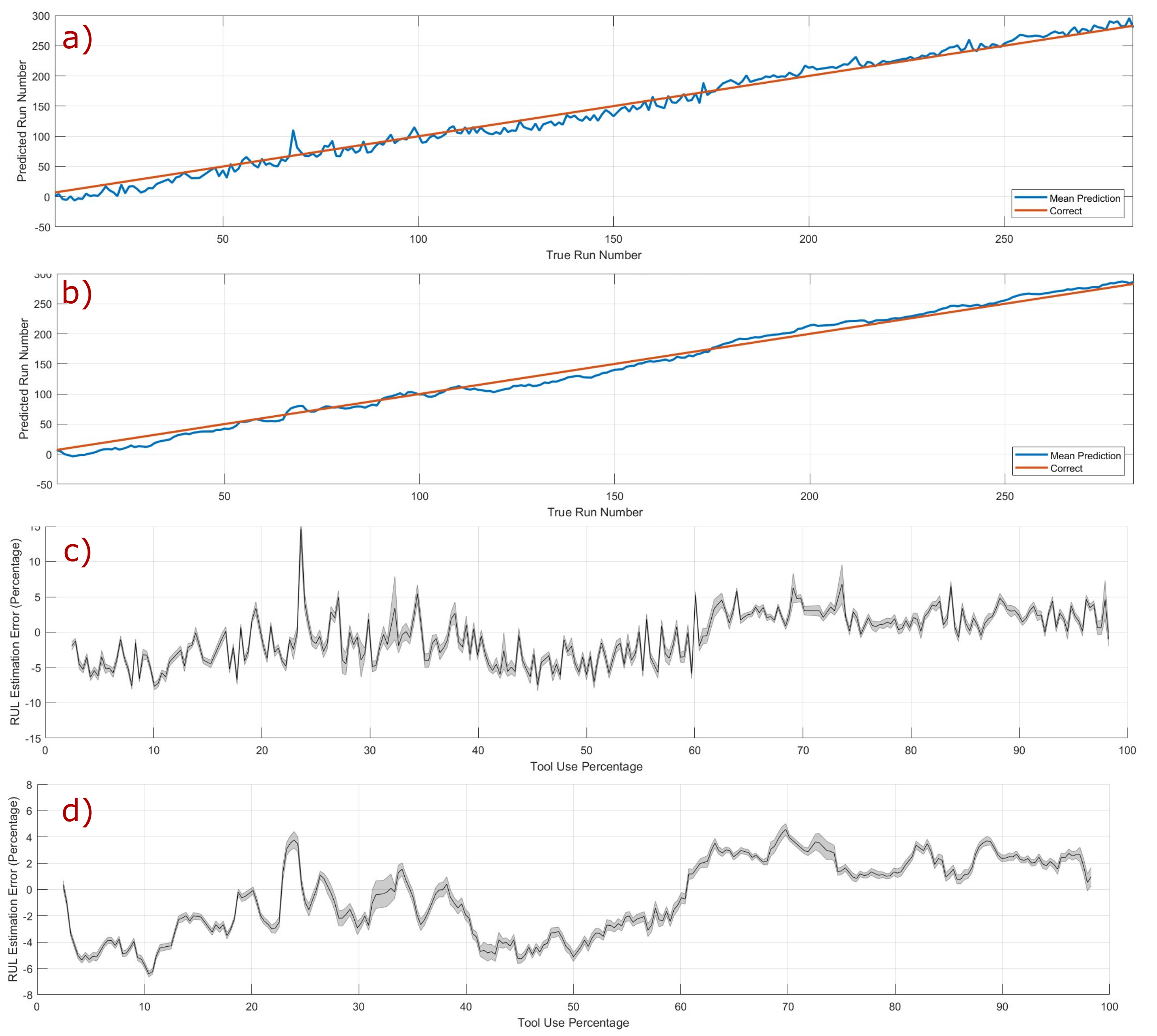}
    \caption{Results of the tool life prediction performed using the CNN networks based on the spectrogram inputs. Panel a) shows the run-number prediction based on a single spectrogram, and panel b) shows the prediction when using the predictions of five spectrograms centered around the desired run number. Panels c) and d) show the mean run number prediction error including the variance bands on the prediction (grey shadows).} 
    \label{fig:results}
\end{figure}

After training we used our test data to evaluate the model's performance, for which the results can be seen in figure \ref{fig:results}. In that figure, panel a) shows the run number prediction based on a single acoustic spectrogram, and panel b) shows the prediction when combining the predictions of five spectrograms centered around the desired run number. Panels c) and d) show the mean run number prediction error including the variance bands on the prediction (grey shadows), in percentage of the total tool life. The maximum error of this prediction is plus or minus 6 percent of the total tool life when using the averaged result. This estimator allows tracking the health of the cutting tool, and allows interventions or continued operation based on the real-world state of tool sharpness. Including a window around the current run number decreases the variance on the estimation and estimation error, as is to be expected.

\begin{figure}[t]
    \centering
    \includegraphics[width=\linewidth]{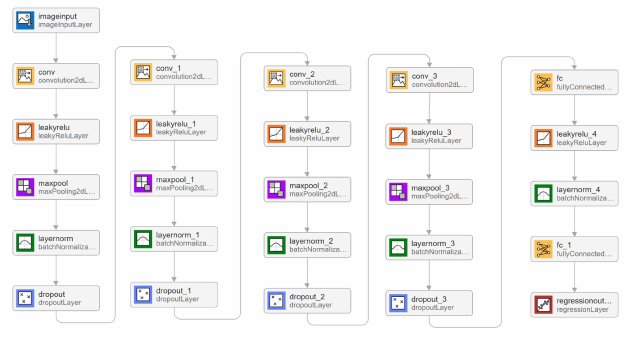}
    \caption{Overview of the CNN architecture, consisting of 4 convolutional layers, followed by a leaky ReLu activation, average pooling, batch normalization and dropout. After the convolutional layers, fully connected layers are used to form the regression outputs.} 
    \label{fig:cnn_overview}
\end{figure}

\section{Discussion and Conclusion}
In this paper, we showed a multi-modal sensor setup for estimating tool wear in CNC turning operations. We discussed the data-acquisition process, and explained what pre-processing steps were taken in data preparation. Subsequently, we detailed the architecture of a convolutional neural network which was applied to the prediction of remaining useful life of the CNC cutting insert. Using an extensive dataset, consisting of 350 tensile test pieces in two materials, we showed the performance of the regression capabilities of these CNNs. We showed that these CNNs are able to estimate the tool life with a maximum error of 6 percent within the total tool life, illustrating the rich data that ultrasonic sensing provides to extract the relevant metrics for predictive maintenance. In future work, we will benchmark our approach to existing approaches in the literature, and evaluate in detail which aspects of the system are relevant: what advantages does a microphone array bring, what benefit has a CNN, and how important is the CNN architecture. 

Furthermore, we will evaluate the problem of generalizing this method to different machines/parts. The system as it is currently proposed, requires a training run for each machine/part/tool combination. In industrial practice, this is not a major hindrance, because each part is often run over multiple weeks, and performing a single training run on an insert is not a large hindrance. However, it would be interesting to implement models which are capable of generalization. For this, larger datasets need to be captured on different machines, making different parts, which is why we classify this as future work.

\clearpage
\newpage

\bibliographystyle{IEEEtran}
\bibliography{references_Zotero}

\end{document}